\documentclass{article}
\usepackage{spconf,amsmath,graphicx}
\usepackage{amssymb}
\usepackage{bm}
\usepackage{xcolor}
\usepackage{qcircuit}
\usepackage{braket}
\usepackage{algorithm}
\usepackage{algorithmicx}
\usepackage{algpseudocode}
\usepackage{mathtools}
\usepackage{hyperref}
\usepackage{setspace}
\usepackage{microtype}

\renewcommand{\figureautorefname}{Figure~\negthinspace}

\renewcommand{\tableautorefname}{Table~\negthinspace}
\renewcommand{\sectionautorefname}{Section~\negthinspace}

\title{Quantum deep recurrent reinforcement learning}
\name{Samuel Yen-Chi Chen \thanks{The views expressed in this article are those of the authors and do not represent the views of Wells Fargo. This article is for informational purposes only. Nothing contained in this article should be construed as investment advice. Wells Fargo makes no express or implied warranties and expressly disclaims all legal, tax, and accounting implications related to this article.}}
\address{Wells Fargo}
\begin{document}
%
\maketitle
\begin{abstract}
Recent advances in quantum computing (QC) and machine learning (ML) have drawn significant attention to the development of quantum machine learning (QML). Reinforcement learning (RL) is one of the ML paradigms which can be used to solve complex sequential decision making problems. Classical RL has been shown to be capable to solve various challenging tasks.
However, RL algorithms in the quantum world are still in their infancy. One of the challenges yet to solve is how to train quantum RL in the partially observable environments. 
In this paper, we approach this challenge through building QRL agents with quantum recurrent neural networks (QRNN). Specifically, we choose the quantum long short-term memory (QLSTM) to be the core of the QRL agent and train the whole model with deep $Q$-learning.
We demonstrate the results via numerical simulations that the QLSTM-DRQN can solve standard benchmark such as Cart-Pole with more stable and higher average scores than classical DRQN with similar architecture and number of model parameters. 
\end{abstract}
\section{Introduction}
\label{sec:intro}
\begin{figure}[htbp]
\centering
\includegraphics[width=0.69\linewidth]{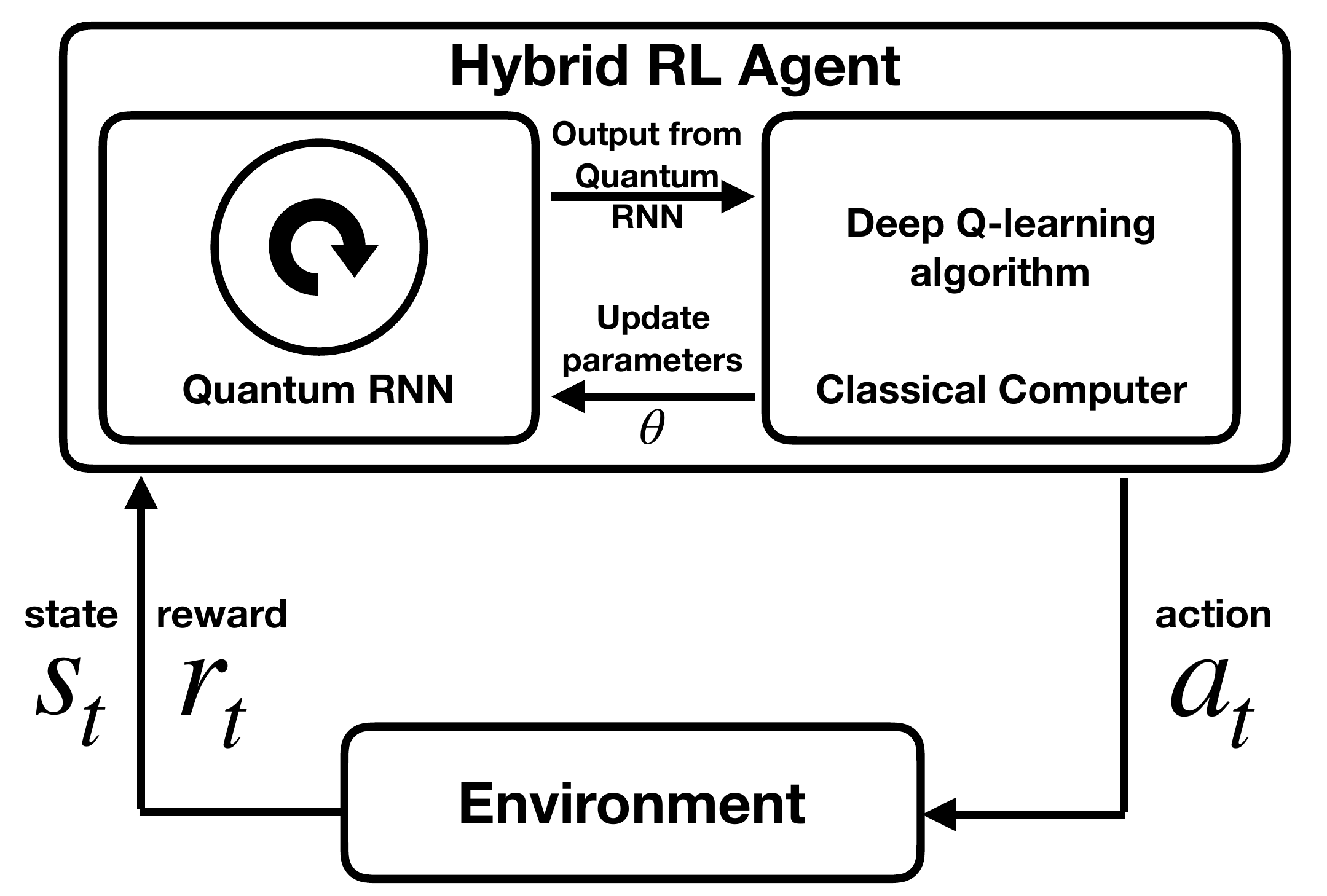}
\caption{{\bfseries The hybrid quantum-classical deep recurrent $Q$-learning.}}
\label{fig:Overall}
\end{figure}

Quantum computing (QC) promises superior performance on certain hard computational tasks over classical computers \cite{nielsen2010quantum}. However, existing quantum computers are not error-corrected, making implementation of deep quantum circuits extremely difficult. These so-called noisy intermediate-scale quantum (NISQ) devices \cite{preskill2018quantum} require special design of quantum circuit architectures so that the quantum advantages can be harnessed.
Recently, a hybrid quantum-classical computing framework \cite{bharti2022noisy} which leverage both the classical and quantum computing has been proposed. Under this paradigm, certain computational tasks which are expected to have quantum advantages are carried out on a quantum computer, while other tasks such as gradient calculations remain on the classical computers.
These algorithms are usually called \emph{variational quantum algorithms} are have been successful in certain ML tasks.
Reinforcement learning (RL) is a sub-field of ML dealing with sequential decision making tasks. RL based on deep neural networks have gained tremendous success in complex tasks with human-level \cite{mnih2015human} or super-human performance \cite{silver2017mastering}. However, quantum RL is an emerging subject with many issues and challenges not yet investigated.
For example, existing quantum RL methods focus on various VQCs without the recurrent structures. Nevertheless, recurrent connections are crucial components in the classical ML to keep memory of past time steps. The potential of such architectures, to our best knowledge, is not yet studied in the quantum RL.
In this work, we propose the quantum deep recurrent $Q$-learning via the application of quantum recurrent neural networks (QRNN) as the value function approximator. Specifically, we apply the quantum long short-term memory (QLSTM) as the core of the QRL agent. The scheme is illustrated in \figureautorefname{\ref{fig:Overall}}.
Our numerical simulation shows that the proposed framework can reach performance comparable to or better than their classical LSTM counterparts when the model sizes are similar and under the same training setting. 
\section{Related Work}
\label{sec:related_work}
The quantum reinforcement learning (QRL) can be traced back to the work \cite{dong2008quantum}. However, the framework requires the environment to be quantum, which may not be satisfied in most real-world cases. Here we focus on the recent developments of VQC-based QRL dealing with classical environments. The first VQC-based QRL \cite{chen19}, which is the quantum version of deep $Q$-learning (DQN), considers discrete observation and action spaces in the testing environments such as Frozen-Lake and Cognitive-Radio. Later, more advanced works in the direction of quantum deep $Q$-learning consider continuous observation spaces such as Cart-Pole \cite{lockwood2020reinforcement,skolik2021quantum}. Hybrid quantum-classical linear solver are also used to find value functions. \cite{Chih-ChiehCHEN2020}. A further improvement of DQN to Double DQN (DDQN) in the VQC framework is considered in the work \cite{heimann2022quantum}, which applies QRL to solve robot navigation task. In addition to learning the \emph{value functions} such as the $Q$-function, QRL frameworks designed to learn the \emph{policy functions} $\pi$ have been proposed recently. For example, the paper \cite{jerbi2021variational} describes the quantum policy gradient RL through the use of REINFORCE algorithm. Then, the work \cite{hsiao2022unentangled} consider an improved policy gradient algorithm called PPO with VQCs and show that quantum models with small amount of parameters can beat their classical counterparts. Along this direction, various modified quantum policy gradient algorithms are proposed such as actor-critic \cite{schenk2022hybrid} and soft actor-critic (SAC) \cite{lan2021variational}. There are also applications of QRL in quantum control \cite{sequeira2022variational} and the proposal of QRL in the multi-agent setting \cite{yun2022quantum}. QRL optimization with evolutionary optimization is first studied in \cite{chen2022variational}. Advanced quantum policy gradient such as DDPG which can deal with continuous action space is considered in \cite{wu2020quantum}. In this work we consider the quantum deep recurrent $Q$-learning, which is an extension of the DQN discussed in \cite{chen19,lockwood2020reinforcement,skolik2021quantum}. However, our work is different from the aforementioned ones as our work consider the \emph{recurrent} quantum policy, while previous works consider the application of various feed-forward VQC architectures. The recurrent policies considered in this work may provide benefits in environments requiring memory of previous steps. 
\section{Reinforcement Learning}
\label{sec:RL}
\emph{Reinforcement learning} (RL) is a machine learning framework in which an \emph{agent} learns to achieve a given goal through interacting with an \emph{environment} $\mathcal{E}$ over a sequence of discrete time steps~\cite{sutton2018reinforcement}. At each time step $t$, the agent observes a \emph{state} $s_t$ and then selects an \emph{action} $a_t$ from the action space $\mathcal{A}$ according to its current \emph{policy} $\pi$. The policy is a mapping from a certain state $s_t$ to the probabilities of selecting an action from $\mathcal{A}$. After exercising the action $a_t$, the agent receives a scalar \emph{reward} $r_t$ and the state of the next time step $s_{t+1}$ from the environment. For episodic tasks, the process proceeds over a number of time steps until the agent reaches the terminal state or the maximum steps allowed. Seeing the state $s_t$ along the training process, the agent aims to maximize the expected return, which can be expressed as the value function at state $s$ under policy $\pi$, $V^\pi(s) = \mathbb{E}\left[R_t|s_t = s\right]$, where $R_t = \sum_{t'=t}^{T} \gamma^{t'-t} r_{t'}$ is the \emph{return}, the total discounted reward from time step $t$. The value function can be further expressed as $V^\pi(s) = \sum_{a\in\mathcal{A}} Q^\pi (s,a) \pi(a|s)$, where the \emph{action-value function} or \emph{Q-value function} $ Q^\pi (s,a) = \mathbb{E}[R_t|s_t = s, a]$ is the expected return of choosing an action $a \in \mathcal{A}$ in state $s$ according to the policy $\pi$. The $Q$-learning is RL algorithm to optimize the $Q^\pi (s,a)$ via the following benchmark formula
\begin{align}
  Q\left(s_{t}, a_{t}\right) \leftarrow & \, Q\left(s_{t}, a_{t}\right)\nonumber\\
  &+\alpha\left[r_{t}+\gamma \max _{a} Q\left(s_{t+1}, a\right)-Q\left(s_{t}, a_{t}\right)\right].
\end{align}
The deep $Q$-learning \cite{mnih2015human} is an extension of the $Q$-learning via the inclusion of the \emph{experience replay} and the \emph{target network} to make the training of deep neural network-based $Q$-value function numerically stable. The deep recurrent $Q$-learning is when a recurrent neural network (RNN) is used to approximate the $Q$-value function \cite{hausknecht2015deep}.
\section{Variational Quantum Circuits}
\label{sec:VQC}
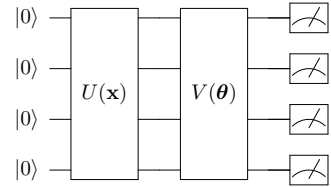
\begin{figure}[hbtp]
\begin{center}
\scalebox{0.8}{
\begin{minipage}{10cm}
\Qcircuit @C=1em @R=1em {
\lstick{\ket{0}} & \multigate{3}{U(\mathbf{x})}  & \qw        & \multigate{3}{V(\boldsymbol{\theta})}       & \qw      & \meter \qw \\
\lstick{\ket{0}} & \ghost{U(\mathbf{x})}         & \qw        & \ghost{V(\boldsymbol{\theta})}              & \qw      & \meter \qw \\
\lstick{\ket{0}} & \ghost{U(\mathbf{x})}         & \qw        & \ghost{V(\boldsymbol{\theta})}              & \qw      & \meter \qw \\
\lstick{\ket{0}} & \ghost{U(\mathbf{x})}         & \qw        & \ghost{V(\boldsymbol{\theta})}              & \qw      & \meter \qw \\
}
\end{minipage}
}
\end{center}
\caption{{\bfseries Generic architecture for variational quantum circuits (VQC).}}
\label{Fig:GeneralVQC}
\end{figure}

Variational quantum circuits (VQC), also known as parameterized quantum circuits (PQC) in the literature, are a special kind of quantum circuits with trainable parameters. Those parameters are trained via optimization algorithms developed in the classical ML communities. The optimization can be gradient-based or gradient-free. The general form of a VQC is described in \figureautorefname{\ref{Fig:GeneralVQC}}
There are three components in a VQC. The \emph{encoding} block $U(\mathbf{x})$ is to transform the classical data $\mathbf{x}$ into a quantum state. The \emph{variational} or \emph{parameterized} block $V(\boldsymbol{\theta})$ represents the part with \emph{learnable} parameters $\boldsymbol{\theta}$ that is optimized through gradient-descent method in this study. The final \emph{measurement} is to output information through measuring a subset (or all) of the qubits and thereby retrieving a (classical) bit string. If we run the circuit once, we can get a bit string such as $0,0,1,1$. However, if we run the circuit multiple times, we can get the expectation values of each qubit. In this paper, we consider the Pauli-$Z$ expectation values of the VQC.

One of the benefits provided by the VQCs is that such circuits are more robust against quantum device noise \cite{kandala2017hardware,farhi2014quantum,mcclean2016theory}. This feature makes VQCs useful for the NISQ \cite{preskill2018quantum} era. In addition, it has been shown that VQCs mav be more expressive than classical neural networks \cite{sim2019expressibility,lanting2014entanglement,du2018expressive,abbas2021power} and can be trained with smaller dataset \cite{caro2022generalization}. Notable examples of VQC in QML include classification \cite{mitarai2018quantum,qi2021qtn,chehimi2022quantum,chen2021federated,chen2021end}, natural language processing \cite{yang2020decentralizing,yang2022bert,di2022dawn} and sequence modeling \cite{chen2020quantum,bausch2020recurrent}.

\section{METHODS}
\label{sec:methods}
\subsection{QLSTM}
\label{sec:qlstm}
\begin{figure}[htbp]
\centering
\includegraphics[width=0.8\linewidth]{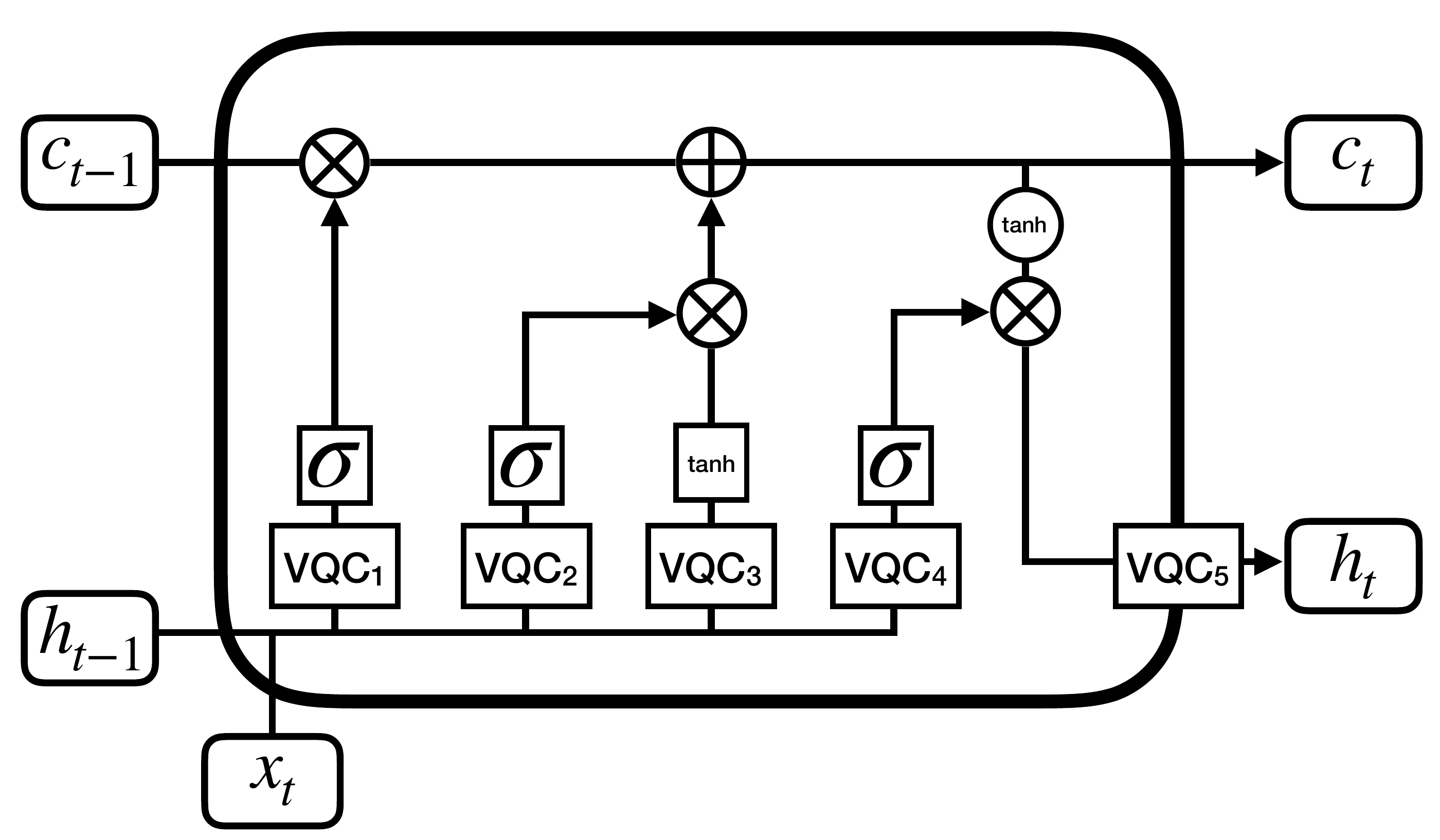}
\caption{{\bfseries The QLSTM used in the proposed quantum deep recurrent $Q$-learning.}}
\label{fig:QLSTM}
\end{figure}
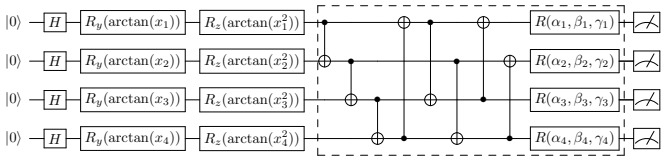
\begin{figure}[htbp]
\begin{center}
\scalebox{0.55}{
\begin{minipage}{10cm}
\Qcircuit @C=1em @R=1em {
\lstick{\ket{0}} & \gate{H} & \gate{R_y(\arctan(x_1))} & \gate{R_z(\arctan(x_1^2))} & \ctrl{1}   & \qw       & \qw      & \targ    & \ctrl{2}   & \qw      & \targ    & \qw      & \gate{R(\alpha_1, \beta_1, \gamma_1)} & \meter \qw \\
\lstick{\ket{0}} & \gate{H} & \gate{R_y(\arctan(x_2))} & \gate{R_z(\arctan(x_2^2))} & \targ      & \ctrl{1}  & \qw      & \qw      & \qw        & \ctrl{2} & \qw      & \targ    & \gate{R(\alpha_2, \beta_2, \gamma_2)} & \meter \qw \\
\lstick{\ket{0}} & \gate{H} & \gate{R_y(\arctan(x_3))} & \gate{R_z(\arctan(x_3^2))} & \qw        & \targ     & \ctrl{1} & \qw      & \targ      & \qw      & \ctrl{-2}& \qw      & \gate{R(\alpha_3, \beta_3, \gamma_3)} & \meter \qw \\
\lstick{\ket{0}} & \gate{H} & \gate{R_y(\arctan(x_4))} & \gate{R_z(\arctan(x_4^2))} & \qw        & \qw       & \targ    & \ctrl{-3}& \qw        & \targ    & \qw      & \ctrl{-2}& \gate{R(\alpha_4, \beta_4, \gamma_4)} & \meter \gategroup{1}{5}{4}{13}{.7em}{--}\qw 
}
\end{minipage}}
\end{center}
\caption{{\bfseries VQC architecture for QLSTM}}
\label{Fig:Basic_VQC_Hadamard_MoreEntangle}
\end{figure}

The QLSTM (shown in \figureautorefname{\ref{fig:QLSTM}}), first proposed in the work \cite{chen2020quantum}, is the quantum version of LSTM \cite{hochreiter1997long}. The main idea behind this model is that the classical neural networks are replaced by the VQCs. It has been shown to be capable of learning time-series data \cite{chen2020quantum} and performing NLP tasks \cite{di2022dawn}.
A formal mathematical formulation of a QLSTM cell is given by
\begin{subequations}
\allowdisplaybreaks
    \begin{align}
    f_{t} &= \sigma\left(VQC_{1}(v_t)\right) \label{eqn:qlstm-f}\\
    i_{t} &= \sigma\left(VQC_{2}(v_t)\right) \label{eqn:qlstm-i}\\ 
    \tilde{C}_{t} &= \tanh \left(VQC_{3}(v_t)\right) \label{eqn:qlstm-bigC}\\
    c_{t} &= f_{t} * c_{t-1} + i_{t} * \tilde{C}_{t} \label{eqn:qlstm-c}\\
    o_{t} &= \sigma\left(VQC_{4}(v_t)\right) \label{eqn:qlstm-o}\\ 
    h_{t} &= VQC_{5}(o_{t} * \tanh \left(c_{t}\right)) \label{eqn:qlstm-h}
    \end{align}
    \label{eqn:qlstm}
\end{subequations}
where the input to the QLSTM is the concatenation $v_t$ of the hidden state $h_{t-1}$ from the previous time step and the current input vector $x_t$ which is the processed observation from the environment. The VQC component used in the QLSTM follows the design shown in \figureautorefname{\ref{Fig:Basic_VQC_Hadamard_MoreEntangle}}. This architecture has been shown to be highly successful in time-series modeling \cite{chen2020quantum}. The data encoding part includes $R_{y}$ and $R_{z}$ rotations, the variational part includes multiple CNOT gates to entangle qubits and trainable general unitary $U$ gates. Finally, the quantum measurement follows.

\subsection{Quantum Deep Recurrent Q-Learning}
\label{sec:QDRQN}
The proposed QDRQN includes the policy network $\theta$ and the target network $\theta^{-}$. Both of them are of the same architecture with a classical NN layer for preprocessing the input, a QLSTM core for the main decision making and a final classical NN layer for post-processing. Such kind of architecture is called \emph{dressed QLSTM model}. During the training, the trajectories are stored in the replay memory so that in the optimization stage, these trajectories can be sampled and make the training more stable. The target network is updated every $S$ steps.
The QDRQN algorithm is summarized in Algorithm. \ref{QDRQN_alg}.
\begin{center}
\scalebox{0.9}{
\begin{minipage}{\linewidth}

\begin{algorithm}[H]
\begin{algorithmic}
\State Initialize replay memory $\mathcal{D}$ to capacity $N$
\State Initialize action-value function dressed QLSTM $Q$ with random parameters $\theta$
\State Initialize target dressed QLSTM $Q$ with $\theta^{-} = \theta$
\For{episode $=1,2,\ldots,M$} 
\State Initialize the episode record buffer $\mathcal{M}$
\State Initialise state $s_1$ and encode into the quantum state
\State Initialize $h_1$ and $c_1$ for the QLSTM
\For {$t=1,2,\ldots,T$}
	\State With probability $\epsilon$ select a random action $a_t$
	\State otherwise select $a_t = \max_{a} Q^*(s_t, a; \theta)$ from the output of the QLSTM
	\State Execute action $a_t$ in emulator and observe reward $r_t$ and next state $s_{t+1}$
	\State Store transition $\left(s_t,a_t,r_t,s_{t+1}\right)$ in $\mathcal{M}$
	\State Sample random batch of trajectories $\mathcal{T}$ from $\mathcal{D}$
	\State Set
	$y_j =
    \left\{
    \begin{array}{l}
      r_j  \quad  \text{for terminal } s_{j+1}\\
      r_j + \gamma \max_{a'} Q(s_{j+1}, a'; \theta) \\  \quad \text{for non-terminal } s_{j+1}
    \end{array} \right.$
	\State Perform a gradient descent step on $\left(y_j - Q(s_j, a_j; \theta^{-}) \right)^2$
        \State Update the target network $\theta^{-}$ every $S$ steps.
\EndFor
\State Store episode record $\mathcal{M}$ to $\mathcal{D}$
\State Update $\epsilon$
\EndFor

\end{algorithmic}
\caption{Quantum deep recurrent $Q$-learning}
\label{QDRQN_alg}
\end{algorithm}
\end{minipage}
}
\end{center}
\section{Experiments}
\label{sec:experiments}
\subsection{Environment}
\label{sec:environment}
The testing environment we choose in this work is Cart-Pole, a standard control problem for benchmarking simple RL models and is also a commonly used example in the OpenAI Gym \cite{brockman2016openai}. In this environment, a pole is attached by a fixed joint to a cart moving horizontally along a frictionless track. The goal of the RL agent is to learn to output the appropriate action according to the observation it receives at each time step so that it can keep the pole as close to the initial state (upright) as possible by pushing the cart leftwards and rightwards. The Cart-Pole environment mapping is: \emph{Observation}: A four dimensional vector $s_t$ including values of the {cart position, cart velocity, pole angle, and pole velocity at the tip}; \emph{Action}: There are two actions $+1$ (pushing rightwards) and $-1$ (pushing leftwards) in the action space; \emph{Reward}: A reward of $+1$ is given for every time step where the pole remains close to being upright. An episode terminates if the pole is angled over $15$ degrees from vertical, or the cart moves away from the center more than $2.4$ units.
\subsection{Hyperparameters}
The hyperparameters for the proposed QDRQN are: batch size: 8, learning rate (Adam): $1 \times 10^{-3}$, memory buffer length $N$: 100, lookup steps $L$: 10.
The $\epsilon$-greedy strategy used in this work is $\epsilon \leftarrow \epsilon \times 0.995$ with the initial $\epsilon = 0.1$ and the final $\epsilon = 0.001$
The target network $\theta^{-}$ is updated every  $S = 4$ steps with the soft update $\theta^{-} \leftarrow \tau \times \theta + (1 - \tau) \times \theta^{-}$ where $\tau = 1\times10^{-2}$.
\subsection{Model Size}
The QLSTM used in this paper includes $8$-qubit VQCs. The input and hidden dimension of the QLSTM are both $4$. The cell or internal state is $8$ dimensional. We consider QLSTM with $1$ or $2$ VQC layers (dashed box in \figureautorefname{\ref{Fig:Basic_VQC_Hadamard_MoreEntangle}}).
We consider the classical LSTM with different number of hidden neurons in the DRQN as the benchmarks. To make fair comparison, we set classical LSTM with model size (number of parameters) similar to the QLSTM model. The LSTM we consider in this work include LSTM with $8$ and $16$ hidden neurons. The summary of quantum and classical models is in \tableautorefname{\ref{tab:number_of_parameters}}.
\begin{table}[htbp]
\begin{tabular}{|l|l|l|l|l|}
\hline
        & QLSTM-1 & QLSTM-2 & LSTM-8 & LSTM-16 \\ \hline
Full    & 150               & 270               & 634           & 2290           \\ \hline
Partial & 146               & 266               & 626           & 2274           \\ \hline
\end{tabular}
\caption{{\bfseries Number of parameters.}}
\label{tab:number_of_parameters}
\end{table}
\subsection{Results}
\begin{figure}[htbp]
\includegraphics[width=0.8\linewidth]{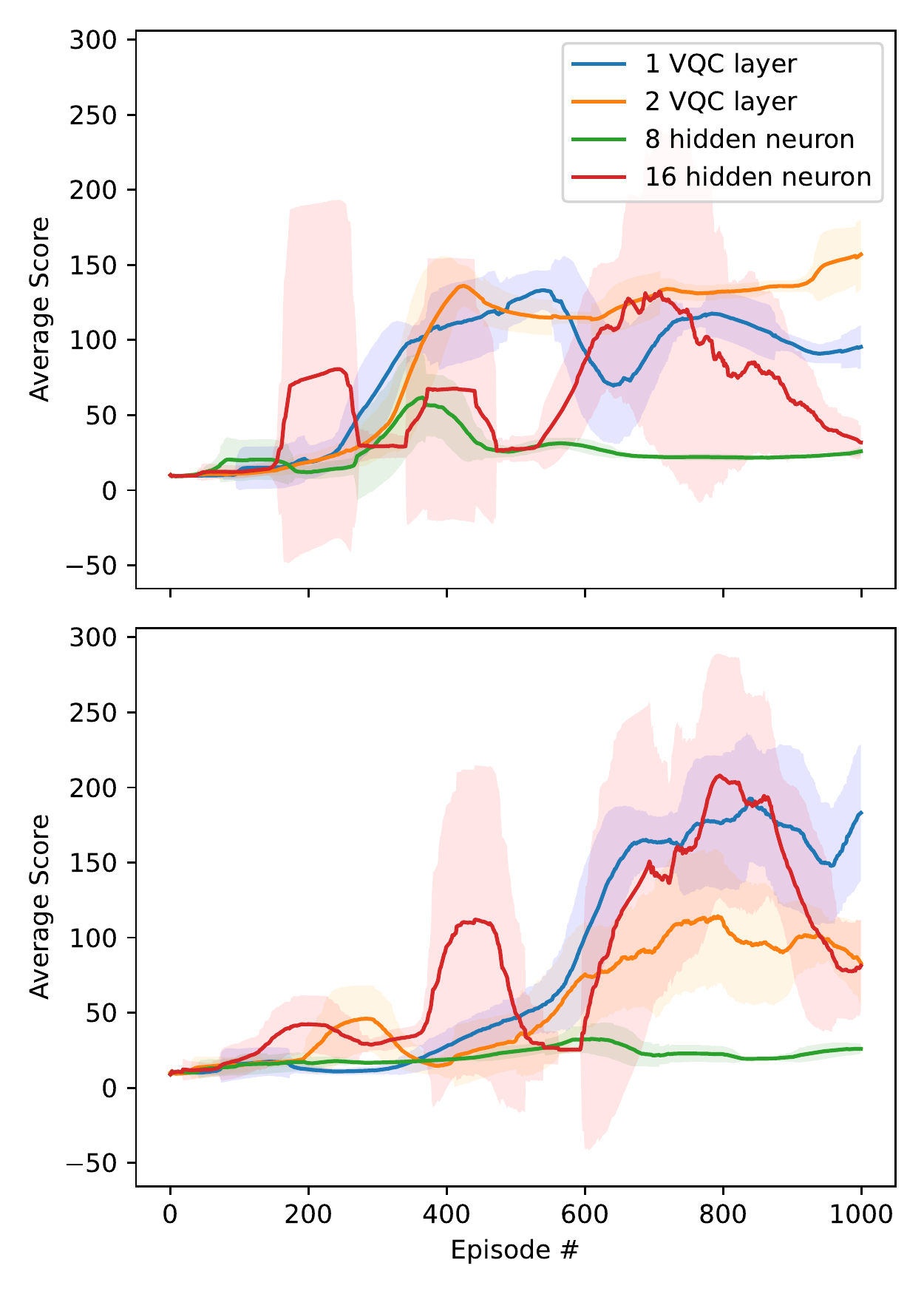}
\caption{{\bfseries Results: Quantum DRQN in Cart-Pole environment.}}
\label{fig:drqn_results}
\end{figure}
\textbf{Fully Observable Cart-Pole} \quad We first consider the standard setting that the RL agent can observe the full state of the environment. In Cart-Pole environment, this is a $4$-dimensional vector includes position and velocities as described in \sectionautorefname{\ref{sec:environment}}. The results are shown in the upper panel of \figureautorefname{\ref{fig:drqn_results}}. We can observe that among the four cases we consider here: QLSTM with 1 or 2 VQC layers and LSTM with 8 or 16 hidden neurons, the QLSTM model with 2 VQC layers achieve the best performance. The QLSTM model with 1 VQC layer performs slightly worse than the one with two VQC layers. We also observe that, under the same training hyperparameters, the quantum models perform much better than their classical counterparts in both the stability and average scores. \\
\textbf{Partially Observable Cart-Pole} \quad We further consider the setting that the RL agent can only observe part of the state of the environment. In Cart-Pole environment, we reduce one dimension of the observation by dropping the last element in the observation vector. Now the RL agent can only get limited information from the environment. The task is more challenging than the previous one since the RL agent needs to figure out how to generate the decision based on partial information. The (Q)LSTM here may be more crucial because the agent may need information from previous steps. We observe that (in the lower panel of \figureautorefname{\ref{fig:drqn_results}}), compared to the fully observed one, in the partially observed environment, the RL agent learns slower (requires more training episodes to reach high scores). Consistent with the fully observed cases, the quantum models perform better than their classical counterparts in both the stability and average scores. For example, the classical LSTM with 16 hidden neurons collapses after 800 episodes of training while the QLSTM agents remain stable.
\section{Conclusions}
\label{sec:conclusions}
In this paper, we first show the quantum recurrent neural network (QRNN)-based RL. Specifically, we employ the QLSTM as the $Q$-function approximator to realize the quantum deep recurrent $Q$-learning. From the results obtained from the testing environments we consider, our proposed framework shows higher stability and average scores than their classical counterparts when the model sizes are similar and training hyperparameters are fixed. The proposed method paves a new way of pursuing QRL with memory of previous time steps.


\clearpage
\begin{spacing}{0.6}
\footnotesize
\bibliographystyle{IEEEbib}
\bibliography{bib/qml_examples,bib/rl,bib/tools,bib/qc_basic,bib/vqc,bib/qrl,bib/classical_ml}

\begin{thebibliography}{10}

\bibitem{nielsen2010quantum}
Michael~A Nielsen and Isaac~L Chuang,
\newblock ``Quantum computation and quantum information,''
\newblock 2010.

\bibitem{preskill2018quantum}
John Preskill,
\newblock ``Quantum computing in the nisq era and beyond,''
\newblock {\em Quantum}, vol. 2, pp. 79, 2018.

\bibitem{bharti2022noisy}
Kishor Bharti, Alba Cervera-Lierta, Thi~Ha Kyaw, Tobias Haug, Sumner
  Alperin-Lea, Abhinav Anand, Matthias Degroote, Hermanni Heimonen, Jakob~S
  Kottmann, Tim Menke, et~al.,
\newblock ``Noisy intermediate-scale quantum algorithms,''
\newblock {\em Reviews of Modern Physics}, vol. 94, no. 1, pp. 015004, 2022.

\bibitem{mnih2015human}
Volodymyr Mnih, Koray Kavukcuoglu, David Silver, Andrei~A Rusu, Joel Veness,
  Marc~G Bellemare, Alex Graves, Martin Riedmiller, Andreas~K Fidjeland, Georg
  Ostrovski, et~al.,
\newblock ``Human-level control through deep reinforcement learning,''
\newblock {\em nature}, vol. 518, no. 7540, pp. 529--533, 2015.

\bibitem{silver2017mastering}
David Silver, Julian Schrittwieser, Karen Simonyan, Ioannis Antonoglou, Aja
  Huang, Arthur Guez, Thomas Hubert, Lucas Baker, Matthew Lai, Adrian Bolton,
  et~al.,
\newblock ``Mastering the game of go without human knowledge,''
\newblock {\em nature}, vol. 550, no. 7676, pp. 354--359, 2017.

\bibitem{dong2008quantum}
Daoyi Dong, Chunlin Chen, Hanxiong Li, and Tzyh-Jong Tarn,
\newblock ``Quantum reinforcement learning,''
\newblock {\em IEEE Transactions on Systems, Man, and Cybernetics, Part B
  (Cybernetics)}, vol. 38, no. 5, pp. 1207--1220, 2008.

\bibitem{chen19}
Samuel Yen-Chi Chen, Chao-Han~Huck Yang, Jun Qi, Pin-Yu Chen, Xiaoli Ma, and
  Hsi-Sheng Goan,
\newblock ``Variational quantum circuits for deep reinforcement learning,''
\newblock {\em IEEE Access}, vol. 8, pp. 141007--141024, 2020.

\bibitem{lockwood2020reinforcement}
Owen Lockwood and Mei Si,
\newblock ``Reinforcement learning with quantum variational circuit,''
\newblock in {\em Proceedings of the AAAI Conference on Artificial Intelligence
  and Interactive Digital Entertainment}, 2020, vol.~16, pp. 245--251.

\bibitem{skolik2021quantum}
Andrea Skolik, Sofiene Jerbi, and Vedran Dunjko,
\newblock ``Quantum agents in the gym: a variational quantum algorithm for deep
  q-learning,''
\newblock {\em Quantum}, vol. 6, pp. 720, 2022.

\bibitem{Chih-ChiehCHEN2020}
Chih-Chieh CHEN, Koudai SHIBA, Masaru SOGABE, Katsuyoshi SAKAMOTO, and Tomah
  SOGABE,
\newblock ``Hybrid quantum-classical ulam-von neumann linear solver-based
  quantum dynamic programing algorithm,''
\newblock {\em Proceedings of the Annual Conference of JSAI}, vol. JSAI2020,
  pp. 2K6ES203--2K6ES203, 2020.

\bibitem{heimann2022quantum}
Dirk Heimann, Hans Hohenfeld, Felix Wiebe, and Frank Kirchner,
\newblock ``Quantum deep reinforcement learning for robot navigation tasks,''
\newblock {\em arXiv preprint arXiv:2202.12180}, 2022.

\bibitem{jerbi2021variational}
Sofiene Jerbi, Casper Gyurik, Simon Marshall, Hans~J Briegel, and Vedran
  Dunjko,
\newblock ``Variational quantum policies for reinforcement learning,''
\newblock {\em arXiv preprint arXiv:2103.05577}, 2021.

\bibitem{hsiao2022unentangled}
Jen-Yueh Hsiao, Yuxuan Du, Wei-Yin Chiang, Min-Hsiu Hsieh, and Hsi-Sheng Goan,
\newblock ``Unentangled quantum reinforcement learning agents in the openai
  gym,''
\newblock {\em arXiv preprint arXiv:2203.14348}, 2022.

\bibitem{schenk2022hybrid}
Michael Schenk, El{\'\i}as~F Combarro, Michele Grossi, Verena Kain, Kevin
  Shing~Bruce Li, Mircea-Marian Popa, and Sofia Vallecorsa,
\newblock ``Hybrid actor-critic algorithm for quantum reinforcement learning at
  cern beam lines,''
\newblock {\em arXiv preprint arXiv:2209.11044}, 2022.

\bibitem{lan2021variational}
Qingfeng Lan,
\newblock ``Variational quantum soft actor-critic,''
\newblock {\em arXiv preprint arXiv:2112.11921}, 2021.

\bibitem{sequeira2022variational}
Andr{\'e} Sequeira, Luis~Paulo Santos, and Lu{\'\i}s~Soares Barbosa,
\newblock ``Variational quantum policy gradients with an application to quantum
  control,''
\newblock {\em arXiv preprint arXiv:2203.10591}, 2022.

\bibitem{yun2022quantum}
Won~Joon Yun, Yunseok Kwak, Jae~Pyoung Kim, Hyunhee Cho, Soyi Jung, Jihong
  Park, and Joongheon Kim,
\newblock ``Quantum multi-agent reinforcement learning via variational quantum
  circuit design,''
\newblock {\em arXiv preprint arXiv:2203.10443}, 2022.

\bibitem{chen2022variational}
Samuel Yen-Chi Chen, Chih-Min Huang, Chia-Wei Hsing, Hsi-Sheng Goan, and
  Ying-Jer Kao,
\newblock ``Variational quantum reinforcement learning via evolutionary
  optimization,''
\newblock {\em Machine Learning: Science and Technology}, vol. 3, no. 1, pp.
  015025, 2022.

\bibitem{wu2020quantum}
Shaojun Wu, Shan Jin, Dingding Wen, and Xiaoting Wang,
\newblock ``Quantum reinforcement learning in continuous action space,''
\newblock {\em arXiv preprint arXiv:2012.10711}, 2020.

\bibitem{sutton2018reinforcement}
Richard~S Sutton and Andrew~G Barto,
\newblock {\em Reinforcement learning: An introduction},
\newblock MIT press, 2018.

\bibitem{hausknecht2015deep}
Matthew Hausknecht and Peter Stone,
\newblock ``Deep recurrent q-learning for partially observable mdps,''
\newblock in {\em 2015 aaai fall symposium series}, 2015.

\bibitem{kandala2017hardware}
Abhinav Kandala, Antonio Mezzacapo, Kristan Temme, Maika Takita, Markus Brink,
  Jerry~M Chow, and Jay~M Gambetta,
\newblock ``Hardware-efficient variational quantum eigensolver for small
  molecules and quantum magnets,''
\newblock {\em Nature}, vol. 549, no. 7671, pp. 242--246, 2017.

\bibitem{farhi2014quantum}
Edward Farhi, Jeffrey Goldstone, and Sam Gutmann,
\newblock ``A quantum approximate optimization algorithm,''
\newblock {\em arXiv preprint arXiv:1411.4028}, 2014.

\bibitem{mcclean2016theory}
Jarrod~R McClean, Jonathan Romero, Ryan Babbush, and Al{\'a}n Aspuru-Guzik,
\newblock ``The theory of variational hybrid quantum-classical algorithms,''
\newblock {\em New Journal of Physics}, vol. 18, no. 2, pp. 023023, 2016.

\bibitem{sim2019expressibility}
Sukin Sim, Peter~D Johnson, and Al{\'a}n Aspuru-Guzik,
\newblock ``Expressibility and entangling capability of parameterized quantum
  circuits for hybrid quantum-classical algorithms,''
\newblock {\em Advanced Quantum Technologies}, vol. 2, no. 12, pp. 1900070,
  2019.

\bibitem{lanting2014entanglement}
Trevor Lanting, Anthony~J Przybysz, A~Yu Smirnov, Federico~M Spedalieri,
  Mohammad~H Amin, Andrew~J Berkley, Richard Harris, Fabio Altomare, Sergio
  Boixo, Paul Bunyk, et~al.,
\newblock ``Entanglement in a quantum annealing processor,''
\newblock {\em Physical Review X}, vol. 4, no. 2, pp. 021041, 2014.

\bibitem{du2018expressive}
Yuxuan Du, Min-Hsiu Hsieh, Tongliang Liu, and Dacheng Tao,
\newblock ``The expressive power of parameterized quantum circuits,''
\newblock {\em arXiv preprint arXiv:1810.11922}, 2018.

\bibitem{abbas2021power}
Amira Abbas, David Sutter, Christa Zoufal, Aur{\'e}lien Lucchi, Alessio
  Figalli, and Stefan Woerner,
\newblock ``The power of quantum neural networks,''
\newblock {\em Nature Computational Science}, vol. 1, no. 6, pp. 403--409,
  2021.

\bibitem{caro2022generalization}
Matthias~C Caro, Hsin-Yuan Huang, Marco Cerezo, Kunal Sharma, Andrew
  Sornborger, Lukasz Cincio, and Patrick~J Coles,
\newblock ``Generalization in quantum machine learning from few training
  data,''
\newblock {\em Nature communications}, vol. 13, no. 1, pp. 1--11, 2022.

\bibitem{mitarai2018quantum}
Kosuke Mitarai, Makoto Negoro, Masahiro Kitagawa, and Keisuke Fujii,
\newblock ``Quantum circuit learning,''
\newblock {\em Physical Review A}, vol. 98, no. 3, pp. 032309, 2018.

\bibitem{qi2021qtn}
Jun Qi, Chao-Han~Huck Yang, and Pin-Yu Chen,
\newblock ``Qtn-vqc: An end-to-end learning framework for quantum neural
  networks,''
\newblock {\em arXiv preprint arXiv:2110.03861}, 2021.

\bibitem{chehimi2022quantum}
Mahdi Chehimi and Walid Saad,
\newblock ``Quantum federated learning with quantum data,''
\newblock in {\em ICASSP 2022-2022 IEEE International Conference on Acoustics,
  Speech and Signal Processing (ICASSP)}. IEEE, 2022, pp. 8617--8621.

\bibitem{chen2021federated}
Samuel Yen-Chi Chen and Shinjae Yoo,
\newblock ``Federated quantum machine learning,''
\newblock {\em Entropy}, vol. 23, no. 4, pp. 460, 2021.

\bibitem{chen2021end}
Samuel Yen-Chi Chen, Chih-Min Huang, Chia-Wei Hsing, and Ying-Jer Kao,
\newblock ``An end-to-end trainable hybrid classical-quantum classifier,''
\newblock {\em Machine Learning: Science and Technology}, vol. 2, no. 4, pp.
  045021, 2021.

\bibitem{yang2020decentralizing}
Chao-Han~Huck Yang, Jun Qi, Samuel Yen-Chi Chen, Pin-Yu Chen, Sabato~Marco
  Siniscalchi, Xiaoli Ma, and Chin-Hui Lee,
\newblock ``Decentralizing feature extraction with quantum convolutional neural
  network for automatic speech recognition,''
\newblock in {\em ICASSP 2021-2021 IEEE International Conference on Acoustics,
  Speech and Signal Processing (ICASSP)}. IEEE, 2021, pp. 6523--6527.

\bibitem{yang2022bert}
Chao-Han~Huck Yang, Jun Qi, Samuel Yen-Chi Chen, Yu~Tsao, and Pin-Yu Chen,
\newblock ``When bert meets quantum temporal convolution learning for text
  classification in heterogeneous computing,''
\newblock in {\em ICASSP 2022-2022 IEEE International Conference on Acoustics,
  Speech and Signal Processing (ICASSP)}. IEEE, 2022, pp. 8602--8606.

\bibitem{di2022dawn}
Riccardo Di~Sipio, Jia-Hong Huang, Samuel Yen-Chi Chen, Stefano Mangini, and
  Marcel Worring,
\newblock ``The dawn of quantum natural language processing,''
\newblock in {\em ICASSP 2022-2022 IEEE International Conference on Acoustics,
  Speech and Signal Processing (ICASSP)}. IEEE, 2022, pp. 8612--8616.

\bibitem{chen2020quantum}
Samuel Yen-Chi Chen, Shinjae Yoo, and Yao-Lung~L Fang,
\newblock ``Quantum long short-term memory,''
\newblock in {\em ICASSP 2022-2022 IEEE International Conference on Acoustics,
  Speech and Signal Processing (ICASSP)}. IEEE, 2022, pp. 8622--8626.

\bibitem{bausch2020recurrent}
Johannes Bausch,
\newblock ``Recurrent quantum neural networks,''
\newblock {\em arXiv preprint arXiv:2006.14619}, 2020.

\bibitem{hochreiter1997long}
Sepp Hochreiter and J{\"u}rgen Schmidhuber,
\newblock ``Long short-term memory,''
\newblock {\em Neural computation}, vol. 9, no. 8, pp. 1735--1780, 1997.

\bibitem{brockman2016openai}
Greg Brockman, Vicki Cheung, Ludwig Pettersson, Jonas Schneider, John Schulman,
  Jie Tang, and Wojciech Zaremba,
\newblock ``Openai gym,''
\newblock {\em arXiv preprint arXiv:1606.01540}, 2016.

\end{thebibliography}
\end{spacing}
\end{document}